\documentclass[aps,twocolumn,showpacs]{revtex4}
\usepackage{graphicx}

\begin{document}

\title{Optically-induced lensing effect on a\\Bose-Einstein condensate expanding in a moving lattice}

\author{L.~Fallani$^{*}$, F.~S.~Cataliotti$^1$, J.~Catani, C.~Fort, M.~Modugno, M.~Zawada$^2$, and M.~Inguscio}

\affiliation{European Laboratory for Non-Linear Spectroscopy (LENS), INFM and
Dipartimento di Fisica, Universit\`a di Firenze, via
Nello Carrara 1, I-50019 Sesto Fiorentino (FI), Italy\\
$^{1}$ also Dipartimento di Fisica, Universit\`a di Catania, via
S. Sofia 64, I-95124 Catania, Italy\\
$^{2}$ also Marian Smoluchowski Physical Institute, Jagiellonian
University, Reymonta 4, 30-059 Krak\'ow, Poland}

\begin{abstract}
We report the experimental observation of a lensing effect on a Bose-Einstein
condensate expanding in a moving 1D optical lattice. The effect of the periodic
potential can be described by an effective mass dependent on the condensate
quasi-momentum. By changing the velocity of the atoms in the frame of the
optical lattice we induce a focusing of the condensate along the lattice
direction. The experimental results are compared with the numerical predictions
of an effective 1D theoretical model. Besides, a precise band spectroscopy of
the system is carried out by looking at the real-space propagation of the
atomic wavepacket in the optical lattice.
\end{abstract}


\date{\today}

\maketitle

The experimental realization of Bose-Einstein condensation allowed
significant advances in the field of atom optics. Forces resulting
from the interaction with coherent light can be used to manipulate
coherent matter waves \cite{atomoptics}. Bragg scattering from two
pulsed laser beams \cite{bragg} has provided a simple tool for the
implementation of atomic mirrors, beam splitters and diffraction
gratings. Superradiant Rayleigh scattering has been used as the
gain mechanism for the development of a coherent matter waves
amplifier \cite{superradiance}. Atom lasers have been realized,
providing pulsed and quasi-cw sources of coherent matter waves
\cite{atomlaser}. Nonlinearities in the atomic wave equation have
been exploited in experiments of nonlinear atom optics, such as
the realization of four wave mixing \cite{fwm} or the observation
of soliton propagation \cite{solitons}. One main difference
between atoms and photons is the mass, which can be modified by
the presence of a periodic potential, such as that resulting from
the interference of two counterpropagating laser beams. The
superfluid behavior of a condensate in such an optical lattice has
been studied in \cite{collective} showing the role of the
effective mass in shifting the collective mode frequencies. The
possibility of achieving experimental control over the effective
mass allows the dispersion management of the matter wavepacket.
Several fascinating effects are predicted to appear in the
negative effective mass regime, such as the formation of gap
solitons in a condensate with repulsive interactions
\cite{gapsolitons}.

In this Letter we demonstrate the possibility to change the
expansion of a Bose-Einstein condensate (BEC) using a moving
optical lattice, which acts as a lens for matter waves, focusing
or defocusing the atomic cloud along the direction of the lattice,
as recently predicted in \cite{massignan02}. The observed
center-of-mass dynamics can be well explained in terms of band
structure and Bloch states, familiar concepts to solid state
physics. This picture has been confirmed by many experimental
results, including the observation of Bloch oscillations
\cite{morsch01} and an extensive work on loading and manipulating
a condensate in an optical lattice \cite{phillips02}. In the rest
frame of the lattice the eigenenergies of the system are $E_n(q)$,
where $q$ is the quasimomentum and $n$ the band index. According
to band theory, the velocity in the $n$-th band is $v_n=\hbar^{-1}
\partial E_n /\partial q$ and the effective mass is $m^*=\hbar^{2}
(\partial^2 E_n /\partial q^2)^{-1}$. We demonstrate that the expansion of the
condensate is strongly modified by the change in the single particle effective
mass $m^*$, which enters the diffusive (kinetic) term in the Gross-Pitaevskii
equation. The expansion of a BEC in a static optical lattice has been already
studied in \cite{morsch02}. Here we use a moving optical lattice to load the
condensate in quasimomentum states with different effective mass. This ability
to access regions of negative effective mass allows us to change the sign of
the matter wave dispersion, inducing the condensate to compress along the
lattice direction instead of expanding \cite{massignan02}.

\begin{figure}
\begin{center}
\includegraphics[width=\columnwidth]{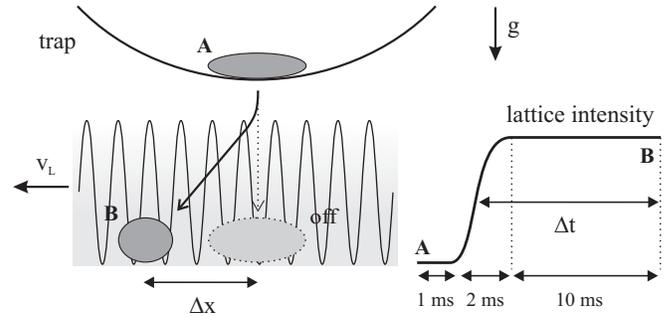}
\caption{Schematics of the experimental procedure. After releasing
the condensate from the magnetic trap (A) we adiabatically ramp
the intensity of an optical lattice moving at velocity $v_L$. We
let the condensate expand in the periodic potential and after 10
ms at the maximum light intensity we look at the position and
dimensions of the atomic cloud by absorption imaging along the
radial horizontal direction (B).} \label{schema}
\end{center}
\end{figure}

The experiment is performed on a Bose-Einstein condensate of
$^{87}$Rb produced in a standard double magneto-optical trap
apparatus by means of combined laser and RF-evaporative cooling.
The evaporation is performed in a Ioffe-Pritchard magnetic trap
with frequencies $\nu_x=8.8$ Hz and $\nu_\perp=90$ Hz along the
axial and radial directions respectively. We typically produce
condensates of $\approx 10^5$ atoms in the hyperfine level
$|F=1,m_F=-1\rangle$ of the ground state. The optical lattice is
provided by two counterpropagating phase-locked laser beams
aligned along the axial direction of the cigar-shaped condensate.
The two beams are circularly polarized and blue-detuned 0.5 nm
from the Rb D2 line at $\lambda=780$ nm. The interference of the
two beams, derived by the same Ti:Sa laser and controlled by two
independent acousto-optic modulators, produces an optical lattice
moving at velocity $v_L=\lambda\Delta\nu/2$, where $\Delta\nu$ is
the frequency difference of the two beams. In the laboratory frame
the resulting potential can be written as $V=sE_R \cos ^2
[k(x-v_Lt)]$, where $k=2\pi/\lambda$ is the modulus of the
wavevector and $s$ measures the depth of the optical lattice in
units of the recoil energy $E_R=\hbar^2k^2/2m$.

The experiment is performed as follows (see Fig. \ref{schema}).
After producing the BEC, we switch off the magnetic trap and let
the atomic cloud expand. After 1 ms of expansion we adiabatically
switch on the moving lattice by ramping the intensity of the two
laser beams in 2 ms. We let the condensate expand in the lattice
and after a total expansion time of 13 ms we take an absorption
image of the cloud along the radial horizontal direction looking
at the position and dimensions of the condensate. We note that the
waist of the laser beams (about 2.0 mm) is big enough to provide a
constant light intensity during the entire expansion of the
condensate. This loading procedure allows us to project the
condensate in a Bloch state of well-defined energy and
quasimomentum \cite{phillips02}. We verified the adiabaticity of
this procedure by checking that, applying the reverse ramp to
switch off the lattice, at the end of the expansion we still have
only one momentum component in the atomic cloud (i.e. we have
populated only one energy band). In our experiments we typically
move the optical lattice with velocities $v_L$ between 0 and
$2v_B$, where $v_B=q_B/m=\hbar k/m = 5.89$ mm/s is the recoil
velocity of an atom absorbing one lattice photon. As a result of
the adiabatic loading we can access different energy bands: for
$0<v_L<v_B$ we populate the first band, while for $v_B<v_L<2v_B$
the second band is populated.

\begin{figure}
\begin{center}
\includegraphics[width=\columnwidth]{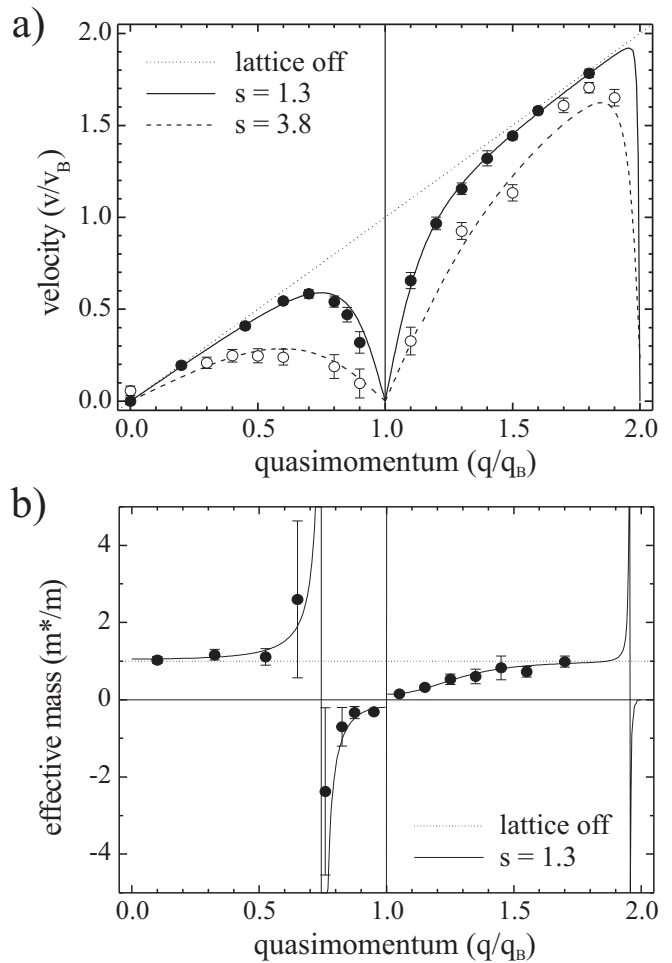}
\caption{a) Velocity of the condensate in the frame of the moving
lattice for the lowest two energy bands and two different optical
intensities: $s=1.3$ (closed circles) and $s=3.8$ (open circles).
The experimental data are obtained from the measured displacements
of the condensate center-of-mass after the expansion inside the
lattice. The lines are calculated from band theory. b) Effective
mass of the condensate in the lowest two energy bands for $s=1.3$.
The experimental points (closed circles) are obtained by
numerically evaluating the incremental ratios $\Delta v/\Delta q$
from the data shown above. The lines are calculated from band
theory. We remember that $v_B=q_B/m=\hbar k/m$.} \label{velmass}
\end{center}
\end{figure}

From the measured positions of the condensate center-of-mass at
the end of the expansion we extract the velocity of the atoms
inside the periodic potential. In the moving frame of the lattice
the atomic velocity is given by $v=v_L-\Delta x/\Delta t$, where
$\Delta x$ is the axial displacement of the condensate and $\Delta
t$ is the time of expansion inside the optical lattice (Fig.~1).
In Fig. \ref{velmass}a we report the experimental velocities as a
function of quasimomentum for two different lattice depths:
$s=1.3(1)$ and $s=3.8(1)$. The error bars include the
indetermination in $\Delta t$ due to the adiabatic switching on of
the optical lattice. The lines shown in the figure are obtained
from the calculation of the velocity in the first two energy
bands. The measured spectrum of velocities shows a very good
agreement with theory. We note that the theoretical curves are
derived from the simple one-particle model neglecting the effect
of interactions. As a matter of fact, since the experiment is
performed after some expansion, we expect that interactions play a
negligible role on the energy spectrum (after 2 ms of expansion
the interaction energy has been almost completely converted into
kinetic energy). An adequate sampling of the experimental
velocities allows us to reconstruct the effective mass by
evaluating the derivative $\partial v/\partial q$ from the finite
increment between consecutive points. In Fig. \ref{velmass}b we
report the results of such analysis on the data taken at $s=1.3$
together with the theoretical curve. This experimental study
allows us to make a precise spectroscopy of the energy bands,
measuring the velocity spectrum and the effective mass of the
condensate in the periodic potential.

\begin{figure}[t]
\begin{center}
\includegraphics[width=\columnwidth]{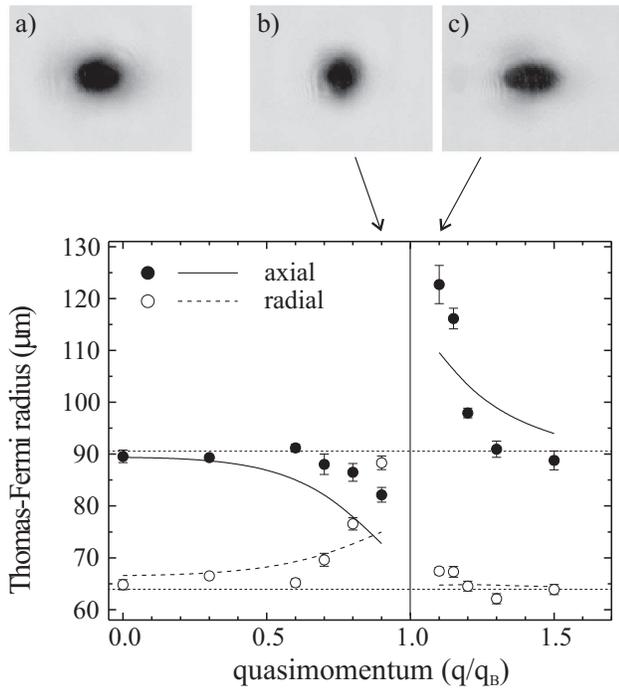}
\caption{Absorption images of the expanded condensate. From left
to right: a) normal expansion of the condensate without lattice;
b) axial compression in a lattice with $s=2.9$ and $v_L=0.9 v_B$;
c) enhanced axial expansion in a lattice with $s=2.9$ and $v_L=1.1
v_B$ (where $v_B=q_B/m=\hbar k/m$). In the lower part we report
the axial and radial dimensions of the condensate after expansion
in an optical lattice with $s=2.9$ as a function of the
quasimomentum. The experimental points (closed and open circles)
show the Thomas-Fermi radii of the cloud extracted from a 2D fit
of the density distribution. The dotted lines show the dimensions
of the expanded condensate in the absence of the optical lattice.
The continuous and dashed lines are theoretical calculations
obtained from the 1D effective model described in the text.}
\label{raggi}
\end{center}
\end{figure}

However, the most interesting feature of this experiment concerns
the observed dependence of the dimensions of the expanded
condensate on different lattice velocities (hence different
quasimomenta of the condensate in the frame of the moving
lattice). Typical absorption images for different $q$ are reported
in the upper part of Fig. \ref{raggi}. In the bottom of Fig.
\ref{raggi} we report the measured Thomas-Fermi radii of the
condensate as obtained from a 2D fit of the measured density
distribution \cite{nota}. We note that, approaching the boundary
of the first Brillouin zone for $q \lesssim q_B$, the axial
dimension of the condensate gets smaller as a consequence of the
modified effective mass $m^*<0$ (see Fig. \ref{raggi}b). In fact,
it is easy to show that a change of sign in the effective mass
corresponds to a time-reversed evolution under the influence of an
inverted external potential (if present). Since in our case the
condensate is initially expanding outwards, when $m^*$ becomes
negative an inversion of dynamics takes place \cite{nota3}. This
contraction continues for times much longer than those considered
in this experiment, until the wavepacket eventually reaches its
minimum allowed size (when dynamics inverts again).

\begin{figure}[b]
\begin{center}
\includegraphics[width=\columnwidth]{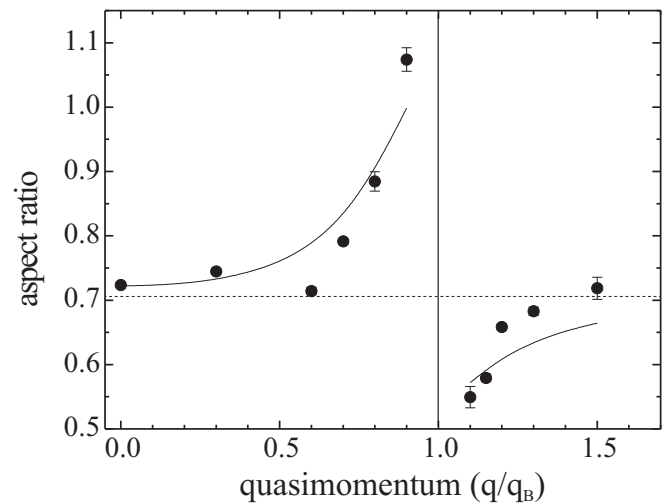}
\caption{Aspect ratio of the condensate after 13 ms of expansion
in an optical lattice with $s=2.9$. The dotted line shows the
aspect ratio of the expanded condensate in the absence of the
optical lattice. The continuous lines is a theoretical calculation
obtained from the 1D effective model described in the text.}
\label{ar}
\end{center}
\end{figure}

This focusing effect along the axial direction is balanced by an
increased expansion along the radial axis, that we attribute to an
effect of interactions. In fact, due to the compression along the
lattice direction, the fast radial expansion is further enhanced
by the increase of the mean-field energy. When the condensate is
loaded in the second band, for $q \gtrsim q_B$, the axial
expansion is enhanced due to the strong positive curvature of the
second energy band near the zone boundary, where $0<m^*<m$ (see
Fig. \ref{raggi}c). As one would expect, in this case the radial
dynamics is not modified, since the residual mean-field energy is
further reduced by the fast axial expansion, causing a suppression
of the non-linear coupling between the axial and radial dynamics.
In Fig. \ref{ar} we also show the (radial to axial) aspect ratio
of the condensate \cite{nota2}, which is characterized by a marked
discontinuity across the boundary between the first and second
zone as predicted in \cite{massignan02}.

To get further insight on the behavior of the condensate during
the expansion, we have analyzed the experimental results by means
of the 1D effective model presented in \cite{massignan02}.
According to this model, the full 3D Gross-Pitaevskii description
of the system is first dynamically rescaled by using the unitary
scaling and gauge transformations of \cite{castin,kagan}, and then
reduced to an effective 1D equation (dr-GPE) by a gaussian
factorization of the radial wavefunction \cite{salasnich}. Despite
its 1D nature, the model is capable to account both for the axial
and radial dynamics of the system, as discussed in
\cite{massignan02}. In the present case the factorization of the
wavefunction is further justified by the fact that during the
expansion the axial and radial degrees of freedom almost decouple.

Actually, Fig. \ref{raggi} shows that the model qualitatively
reproduces the behavior observed in the experiment, even though it
does not fit precisely the data. In particular, approaching the
first zone boundary, the observed focusing effect along the axial
direction is slightly smaller than the calculated one, and at the
same time the expansion along the radial direction (not directly
affected by the lattice) is enhanced. Instead, in the second band
the radial behavior is well reproduced by the model, whereas there
is still a discrepancy concerning the axial expansion. We remark
that in the region near to the band edge ($0.95q_B<q<1.05q_B$) the
process of switching on the optical lattice is no longer
adiabatic, and the description is complicated by the fact that
more than one energy band gets populated. Indeed what we actually
see in the experiment is a superposition of two atomic clouds with
different shapes resulting from the minor population of a
different energy band. At the present stage of the experiment, we
cannot increase arbitrarily the ramp time of the lattice intensity
as the condensate, under the effect of gravity, falls out of the
lattice beams.

In conclusion, we have achieved a lensing effect on a
Bose-Einstein condensate expanding inside a moving optical
lattice. Tuning the velocity of the lattice we can set the lensing
power of the periodic potential, all the way from focusing to
defocusing of the atomic cloud. The demonstrated experimental
control of the matter wave dispersion is likely to open new
possibilities in the field of atom optics, including the
observation of non linear effects such as the generation of gap
solitons. The experimental techniques introduced in this work will
also allow future high precision studies of the band structure of
a Bose-Einstein condensate in an optical lattice.

We acknowledge P. Massignan for stimulating discussions and
theoretical support in the initial stage of this experiment. This
work has been supported by the EU under Contracts Numbers HPRI-CT
1999-00111 and HPRN-CT-2000-00125, by the MURST through the
PRIN2000 Initiatives and by the INFM Progetto di Ricerca Avanzata
``Photon Matter''. M.Z. acknowledges partial support of the grant
PBZ/KBN/043/PO3/2001.

\textit{Note added.} Only after completion of this work we became
aware of the publication of a paper \cite{oberthaler} showing
results similar to those presented in this Letter.


\begin{thebibliography}{99}
\bibitem{nota}
Due to fluctuations ($\sim 20$\%) in the number of atoms in the
different experiments, we scaled the measured radii to the average
number of atoms using the relation $R \propto N^{1/5}$ valid in
the Thomas-Fermi limit. Each data point in Fig.~\ref{raggi}
corresponds to an average of 10 repeated measurements.
\bibitem{nota3}
Note that if the sign of the effective mass $m^*$ changes when the
condensate is still at rest, after the release from the trap it
would expand regardless of the sign of $m^*$.
\bibitem{nota2}
We note that, while the dimensions of the condensate really depends on the
number of atoms, in the Thomas-Fermi limit the aspect ratio is not affected by
fluctuations in the size of the condensate.
\end{thebibliography}
\end{document}